\documentclass[11pt,a4paper]{article}
\pdfoutput=1
\usepackage{graphicx}
\usepackage{jheppub}
\usepackage{cases}
\usepackage{bm}
\usepackage{epstopdf}
\usepackage{epsfig}
\usepackage{enumerate}
\usepackage{color,framed}
\usepackage{amsmath,amsthm,amssymb} 
\usepackage[normalem]{ulem}

\title{Revisit on two-dimensional self-gravitating kinks: superpotential formalism and linear stability}

\author[]{Yuan Zhong\footnote{Corresponding author.}}

\affiliation[]{School of Physics, Xi'an Jiaotong University, Xi'an 710049, People's Republic of China}
 \emailAdd{zhongy@mail.xjtu.edu.cn}

\arxivnumber{2101.10928}

\abstract{Self-gravitating kink solutions of a two-dimensional dilaton gravity are revisited in this work.  Analytical kink solutions are derived from a concise superpotential formalism of the dynamical equations.  A general analysis on the linear stability is conducted for an arbitrary static solution of the model. After gauge fixing, a  Schr\"odinger-like equation with factorizable Hamiltonian operator is obtained, which ensures the linear stability of the solution.  }


\begin{document}
\maketitle  

\section{Introduction}
During the past decades, two-dimensional (2D) gravitational  models continue attracting the attention of theorists for a variety of reasons. First of all, the field equations obtained in many 2D gravity models are simple enough to allow a rigorous analysis of some difficult issues of gravitational theory, such as the quantization of gravity~\cite{Henneaux1985,Alwis1992}, gravitational collapse~\cite{VazWitten1994,VazWitten1996}, black hole evaporation~\cite{CallanGiddingsHarveyStrominger1992,BilalCallan1993,RussoSusskindThorlacius1992,RussoSusskindThorlacius1992a,RussoSusskindThorlacius1993}, see~\cite{Brown1988,Thorlacius1995,GrumillerKummerVassilevich2002} for comprehensive reviews on early works.  Second, a number of very different approaches of quantum gravity all hint that at very short distances space-time becomes effectively two dimensional~\cite{AmbjornJurkiewiczLoll2005,Horava2009b,MureikaStojkovic2011,AnchordoquiDaiFairbairnLandsbergEtAl2012,Stojkovic2013,Loll2020}. Here, the dimensions that are reduced can be effective, spectral, topological or the usual dimensions~\cite{Carlip2017}. Recently, the studies of the Sachdev-Ye-Kitaev (SYK) model \cite{SachdevYe1993,Kitaev2015} also lead to a resurgence of interest in 2D gravity~\cite{AlmheiriPolchinski2015,MaldacenaStanfordYang2016,MaldacenaStanford2016,Jensen2016}, see~\cite{Rosenhaus2018,Sarosi2018,Trunin2020} for pedagogical introductions.

Since the Einstein tensor vanishes identically in two dimensions, the Einstein-Hilbert action cannot be used to describe 2D gravity. An economical solution to this problem is to introduce a dilaton field. Many different 2D dilaton gravity models have been proposed and studied so far. The simplest action for 2D dilaton gravity is the Jackiw-Teitelboim (JT) action~\cite{Jackiw1985,Teitelboim1983}
\begin{eqnarray}
S_{J T}=\frac{1}{\kappa} \int d^{2} x  \sqrt{-g} \varphi(R+\Lambda),
\end{eqnarray}
where the dilaton $\varphi$ plays the role of a Lagrangian multiplier. $\kappa$ and $\Lambda$ are the gravitational coupling and the cosmological constant, respectively.  Two other famous actions for 2D dilaton gravity are the Mann-Morsink-Sikkema-Steele (MMSS) action, which generalize the JT action by giving the dilaton a kinetic term~\cite{MannMorsinkSikkemaSteele1991}
\begin{eqnarray}
\label{MMSSgra}
S_{\textrm{MMSS}}=\frac{1}{\kappa}\int{d^2x}\sqrt{-g}\left[ -\frac{1}{2}(\nabla\varphi)^2  +\varphi  R  +\Lambda\right],
\end{eqnarray}
and the Callan-Giddings-Harvey-Strominger (CGHS) action~\cite{CallanGiddingsHarveyStrominger1992}:
\begin{eqnarray}
S_{\mathrm{CGHS}}=\frac{1}{2\pi} \int d^{2} x \sqrt{-g}\left\{e^{-2 \varphi}\left[R+4 (\nabla\varphi)^2  +4\Lambda^2 \right]-\frac12(\nabla \phi )^2\right\},
\end{eqnarray}
where $\phi$ is a massless scalar matter field.
A comprehensive review of 2D dilaton gravity models and their applications in black hole physics and quantum gravity can be found in Ref. \cite{GrumillerKummerVassilevich2002}.  

It is a natural idea to extend the discussion on 2D dilaton gravity to other classical solutions such as topological solitons, which could be produced by cosmic phase transitions~\cite{VilenkinShellard2000}. As the simplest topological soliton solution, kink (or domain wall) has been extensively studied in 4D cosmology~\cite{Vachaspati2006} and 5D thick brane world models~\cite{DzhunushalievFolomeevMinamitsuji2010,Liu2018}. In the case of two dimensions, previous works have revealed close connections between kinks and 2D black holes~\cite{ShinSoh1995,JohngShinSoh1996,GegenbergKunstatter1998,Cadoni1998}, or naked singularities~\cite{VazWitten1994,VazWitten1996,VazWitten1995,YanQiu1998,YanWangTao2001}. 

In 1995, an exact 2D self-gravitating sine-Gordon kink solution without curvature singularity was found by St\"otzel, in the MMSS gravity model~\cite{Stoetzel1995}. In addition to the kink configuration of the scalar field, the metric solution~\cite{Stoetzel1995} describes a 2D asymptotic anti de-Sitter (AdS$_2$) geometry. This property reminds us the thick brane solutions found in asymptotic AdS$_5$ geometry~\cite{SkenderisTownsend1999,DeWolfeFreedmanGubserKarch2000,Gremm2000}. 
The aim of the present work is to reveal similarities between 2D self-gravitating kinks and 5D thick brane worlds.

The organization of the paper is as follows. In Sec.~\ref{SecModel}, we give a brief review of St\"otzel's model, and show that for static solutions, the field equations can be written as a group of first-order differential equations by introducing the so called superpotential. With the superpotential formalism, one can easily generate exact self-gravitating kink solutions by chosen proper superpotentials. We will discuss two analytical solutions in Sec.~\ref{SecSolution}. Then, in Sec.~\ref{SecSability} we give a complete analysis to the linear stability of the solutions. To our knowledge, no such analysis was done before. In a recent work~\cite{IzquierdoFuertesGuilarte2020}, the authors considered the linear perturbations around self-gravitating kink solutions in 2D MMSS gravity. However, they expand the metric around the Minkowski metric rather  than the asymptotic AdS$_2$ metric solution. Finally, we offer in Sec.~\ref{SecConc} some concluding remarks.

\section{The model and the superpotential formalism}
\label{SecModel}

The action of St\"otzel's model~\cite{Stoetzel1995} contains an MMSS gravity part along with a canonical real scalar $\phi$:
\begin{eqnarray}
\label{action}
S=\frac{1}{\kappa}\int{d^2x}\sqrt{-g}\left[ -\frac{1}{2}\partial ^{\mu}\varphi \partial _{\mu}\varphi +\varphi  R  +\Lambda  +\kappa \mathcal{L}_{\text{m}}\right],
\end{eqnarray}
where
\begin{eqnarray}
\mathcal{L}_{\text{m}}=-\frac{1}{2}\partial ^{\mu}\phi \partial _{\mu}\phi -V(\phi)
\end{eqnarray}
is the Lagrangian density of the scalar field.

After variation, one immediately obtains the Einstein equations
\begin{eqnarray}
\label{eqEinstein}
\nabla _{\mu}\varphi \nabla _{\nu}\varphi +2\nabla _{\mu}\nabla _{\nu}\varphi -\frac{1}{2}g_{\mu \nu}\left( \nabla _{\lambda}\varphi \nabla ^{\lambda}\varphi +4\nabla _{\lambda}\nabla ^{\lambda}\varphi -2\Lambda\right) =-\kappa T_{\mu \nu},
\end{eqnarray}	
the dilaton equation
\begin{eqnarray}
\label{eqDilatonxcoord}
\nabla _{\lambda}\nabla ^{\lambda}\varphi +R =0,
\end{eqnarray}
and the scalar field equation
\begin{eqnarray}
\label{EqEOM}
\nabla ^{\mu}\nabla _{\mu}\phi -\frac{dV}{d\phi}=0.
\end{eqnarray}
The energy-momentum tensor in Eq.~\eqref{eqEinstein} is defined as 
\begin{eqnarray}
T_{\mu \nu}&=&g_{\mu \nu}\mathcal{L}_{\text{m}}-2\frac{\delta \mathcal{L}_{\text{m}}}{\delta g^{\mu \nu}}\nonumber\\
	&=&\partial _{\mu}\phi \partial _{\nu}\phi -\frac{1}{2}g_{\mu \nu}\left(\partial ^{\alpha}\phi \partial _{\alpha}\phi +2V \right).
\end{eqnarray}

To obtain self-gravitating kink solution, St\"otzel used the following metric
\begin{eqnarray}
\label{metricX}
ds^2=-e^{2A(x)}dt^2+dx^2.
\end{eqnarray}
Similar metric ansatz is also used in 5D brane world models with non-factorizable geometry~\cite{RandallSundrum1999a,RandallSundrum1999}, therefore, we will follow the terminology of brane world theory and call the function $A(x)$ as the warp factor. As a convention, the derivative with respect to $x$ will always be denoted as a subscript, for example, $\phi_x\equiv d\phi/dx.$

Substituting metric \eqref{metricX} into the Einstein equations \eqref{eqEinstein}, one obtains
\begin{eqnarray}
\label{EinEq1}
2 A_x  \varphi_x -2 \varphi_{xx}- \varphi_x^2&=&\kappa  \phi _x^2,\\
\label{EinEq2}
 A_x  \varphi_x  + \varphi_{xx} &=& \Lambda- \kappa  V.
\end{eqnarray}
The equations of motion for the dilaton and the scalar fields read
\begin{eqnarray}
\label{Req}
-2 A_{xx}-2 A_{x}^2+\varphi_{xx}+A_{x} \varphi_{x}=0.
\end{eqnarray}
and 
\begin{eqnarray}
\label{EoM}
A_{x} \phi_{x}+\phi_{xx}=\frac{dV}{d\phi},
\end{eqnarray}
respectively.  Note that only three of the above equations are independent. For example, Eq.~\eqref{EoM} can be derived  by using Eqs. \eqref{EinEq1}-\eqref{Req}. 
At a first glance, Eqs. \eqref{EinEq1}-\eqref{EoM} constitute a complicate nonlinear differential system, and finding their solutions seems to be a formidable task. But the study of brane world models has taught us a lesson on how to solve such system by means of superpotential method, which rewrites second-order differential equations, such as Eqs. \eqref{EinEq1}-\eqref{EoM}, into some first-order ones~\cite{SkenderisTownsend1999,DeWolfeFreedmanGubserKarch2000,Gremm2000}.

To construct a superpotential formalism for the present model, we first note that the combination of Eqs. \eqref{EinEq2} and \eqref{Req} leads to an expression of $V$ in terms of cosmological constant and warp factor:
\begin{eqnarray}
\label{eqV}
\kappa V=\Lambda-2A_{xx}-2A_{x}^2.
\end{eqnarray}
Taking the derivative of the above equation and eliminating $dV/d\phi$ by using Eq. \eqref{EoM}, one obtains a relation between $A$ and $\phi$:
\begin{eqnarray}
\label{EqAphi}
A_{xxx}+2A_{x} A_{xx}=-\frac{1}{2}\kappa( A_{x}\phi_{x}^2+ \phi_{xx} \phi_{x}).
\end{eqnarray}
The superpotential method starts with an assumption that the first-order derivative of $\phi$ equals to a function of $\phi$ itself, namely, the superpotential $W(\phi)$ via the following equation:
\begin{eqnarray}
\label{EqPhiW}
\phi_{x}=\frac{dW}{d\phi}.
\end{eqnarray}
Under this assumption, one can testify that Eq. \eqref{EqAphi} supports a very simple special solution:
\begin{eqnarray}
\label{EqAW}
A_{x}=-\frac14 \kappa W.
\end{eqnarray}
Then, Eq. \eqref{eqV} enables us to write $V$ in terms of superpotential:
\begin{eqnarray}
\label{EqVW}
V=\frac{1}{2}\left(\frac{dW}{d\phi}\right)^2-\frac{1}{8}\kappa W^2+\frac{\Lambda}{\kappa}.
\end{eqnarray}
Finally, the general solution of Eq.~\eqref{Req} gives a simple relation between dilaton and warp factor:
\begin{eqnarray}
\varphi=2A+\beta \int e^{-A} dx+\varphi_0, \nonumber
\end{eqnarray}
where $\beta$ and $\varphi_0$ are just two integral constants. Since the field equations only contain the derivatives of the dilaton, the value of $\varphi_0$ is unimportant to the solution of other variables, and can be taken as $\varphi_0=0$.
Besides, to consist with Eq. \eqref{EinEq1}, $\beta$ must be set as zero, so
\begin{eqnarray}
\label{eqAVarphi}
\varphi=2A. 
\end{eqnarray}

Eqs.~\eqref{EqPhiW}-\eqref{eqAVarphi} constitute the first-order superpotential formalism of the present model. Exact kink solutions can be derived by choosing proper superpotentials. The freedom of choosing a superpotential comes from the fact that there are four unknown variables ($A, \phi, \varphi$ and $V$) but only three independent equations. Taking a superpotential amounts to specifying one of the four unknown variables.

\section{Exact solutions}
\label{SecSolution}
In this section, we show how to use the superpotential formalism to derive exact self-gravitating kink solutions. We first reproduce St\"otzel's solution and then report a new solution. 

\subsection{Reproducing St\"otzel's solution}
In fact, the superpotential formalism presented in last section has been derived and used, although unconsciously,  by St\"otzel~\cite{Stoetzel1995}. Instead of choosing a superpotential $W(\phi)$, St\"otzel started with the Sine-Gordon potential
\begin{eqnarray}
V(\phi )=2m^2\sin ^2\frac{\phi}{2}.
\end{eqnarray}
He observed that when $\kappa= \frac{\lambda }{4 m^2-\lambda }$, Eq. \eqref{EqVW} surports two solutions of the superpotential:
\begin{eqnarray}
\label{SolW}
W_{\pm}=\pm 2 \sqrt{4 m^2-\lambda } \cos \left(\frac{\phi }{2}\right),
\end{eqnarray}
where $0<\lambda\equiv \frac{2\Lambda}{\kappa}<4 m^2$. The solutions of $\phi(x)$ corresponds to $W_{-}$ could be obtained by integrating Eqs. \eqref{EqPhiW}, and the result turns out to be the sine-Gordon kink~\cite{Stoetzel1995}:
\begin{eqnarray}
\label{SolPhi}
\phi_K(x)=4 \arctan\left( e^{ M (x-x_0)}\right).
\end{eqnarray}
Here $x_0$ is an integral constant that represents the position of the kink, and will be set to zero from now on. The constant $M$ is defined as $M\equiv \frac{1}{2} \sqrt{4 m^2-\lambda }$. Obviously, $M\in(0,m)$. 
The solution corresponds to $W_{+}$ is an antikink
\begin{eqnarray}
\phi_{\bar{K}}(x)=4 \arctan\left( e^{-M x}\right),
\end{eqnarray} 
which is similar as the kink in many aspects. Thus, we will focus on the kink solution only, and eliminate the subscript $K$ from now on. 

\begin{figure}[h]
\centering
\includegraphics[width=1\textwidth]{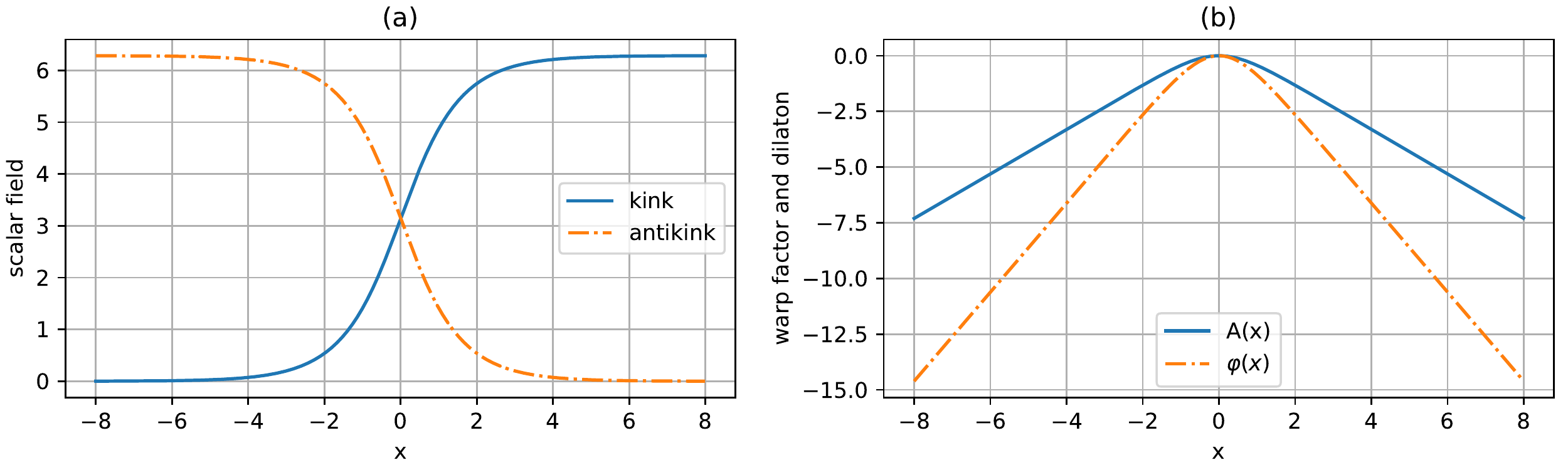}
\caption{The shapes of some important variables in St\"otzel's solution, incluting (a) scalar field, (b) warp factor and the dilaton field. The parameters are taken as $\kappa=1$, $m=\sqrt{2}$ and $\lambda=4$, therefore $M=1$ and $\Lambda=2$. }
\label{figSine}
\end{figure}

Plugging the solutions of  $W(\phi)$ and $\phi(x)$ into Eq.~\eqref{EqAW}, one immediately obtains the expression of the warp factor:
\begin{eqnarray}
A(x)=A_0-\frac{\lambda}{4 M^2}\ln (2 \cosh (M x)),
\end{eqnarray}
which further reduces to~\cite{Stoetzel1995}  
\begin{eqnarray}
A(x)&=&-\frac{\lambda}{4 M^2}\ln \cosh (M x)\nonumber\\
&=&-\kappa\ln \cosh (M x)
\end{eqnarray}
after taking integral constant $A_0=\frac{\lambda}{4 M^2}\ln 2$. Obviously, this warp factor describes an asymptotic AdS$_2$ geometry. Finally, the dilaton field reads
\begin{eqnarray}
\varphi(x)&=&2A(x)=-2\kappa\ln \cosh (M x).
\end{eqnarray}
The profiles  of $\phi$, $A$ and $\varphi$ are plotted in Fig.~\ref{figSine}.

\subsection{A polynomial superpotential}
As shown repeatedly in the study of 5D thick brane models,  it is quite easy to construct exact self-gravitating kink solutions once the superpotential formalism is established. In the following discussions, we will take $\Lambda=0$ for simplicity, as it can be absorbed into the definition of $V(\phi)$.

\begin{figure}[h]
\centering
\includegraphics[width=0.5\textwidth]{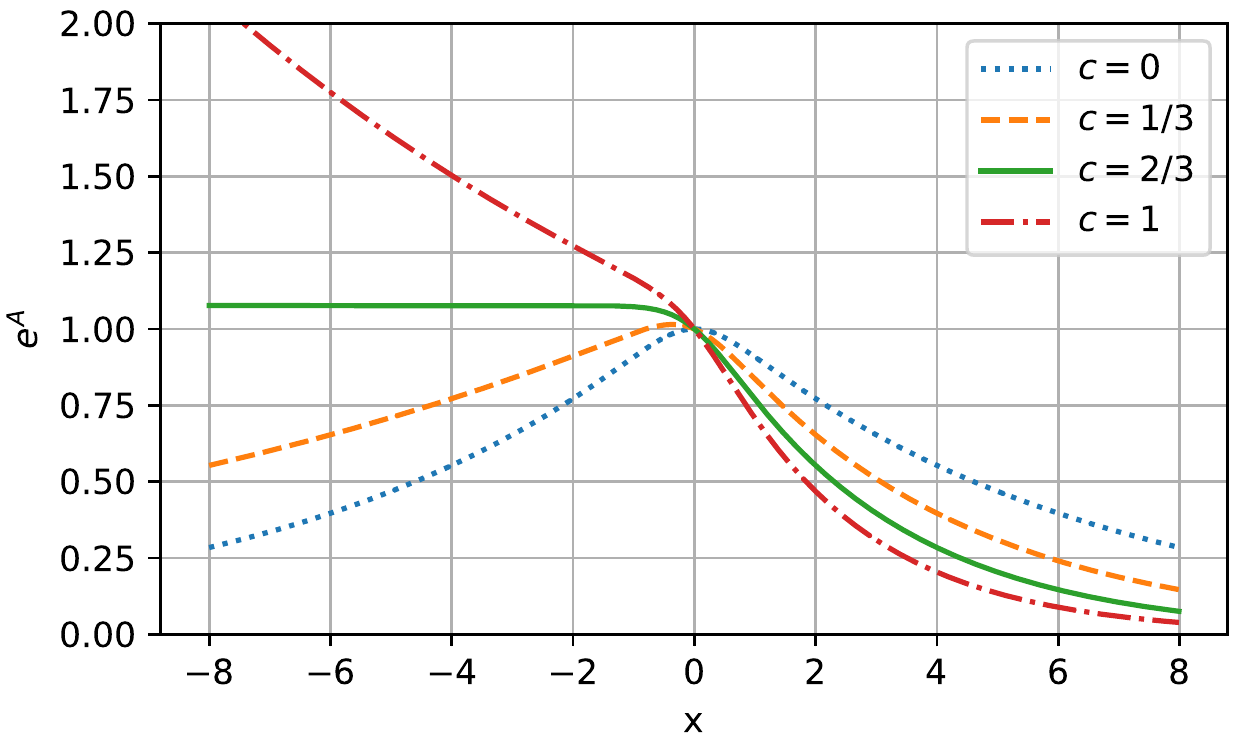}
\caption{Plots of warp factor $e^{A(x)}$ of the polynomial model with $\kappa=1$.}
\label{FigPolyWarp}
\end{figure}

Consider a simple polynomial potential with parameter $c$ \cite{EtoSakai2003,TakamizuMaeda2006,BazeiaMenezesRocha2014}
\begin{eqnarray}
W=c+\phi  \left(1-\frac{\phi ^2}{3}\right).
\end{eqnarray}
It  has two minima at $\phi_{\pm}=\pm1$,  where $W(\phi_{\pm})=\pm\frac23 +c$. With this superpotential, one obtains~\cite{BazeiaMenezesRocha2014}
\begin{eqnarray}
\phi (x)&= & \tanh (x),\\
\varphi(x)&=&2A(x),\\
A(x)&=&\frac{1}{24} \kappa  \left[-6 c x+\text{sech}^2(x)-4 \ln (\cosh (x))-1\right],\\
V(\phi)&=&-\frac{1}{72} \kappa  \left(-3 c+\phi ^3-3 \phi \right)^2+\frac{1}{2} \left(\phi ^2-1\right)^2.
\end{eqnarray}
The asymptotic behaviors of the warp factor and the scalar potential are
\begin{eqnarray}
A_\pm(x)&=&-\frac14 \kappa W(\phi_\pm) x=-\frac14 \kappa (\frac23\pm c)|x|,\\
V_\pm&=&-\frac{1}{72} (3 c\pm2)^2 \kappa.
\end{eqnarray}
Depending on the value of $c$, there are four different situations~\cite{BazeiaMenezesRocha2014}:
\begin{enumerate}
\item $c=0$: In this case, the kink connects two equivalent AdS$_2$ spaces symmetrically, and $V_+=V_-=-\frac{1}{18} \kappa$.
\item $0<|c|<\frac23$: The kink connects two distinct AdS$_2$ spaces.
\item $|c|=\frac23$: The kink connects an AdS$_2$ space and a 2D Minkowski space (M$_2$) asymmetrically. This situation is of particular interesting when considering kink collision in asymptotical AdS space-time~\cite{TakamizuMaeda2006,OmotaniSaffinLouko2011}.
\item $|c|>\frac23$: The warp factor  diverges at one side of the kink.
\end{enumerate}
The behavior of $e^{A}$ for different values of $c$ has been plotted in Fig.~\ref{FigPolyWarp}. Obviously, for $c\neq 0$, the warp factor is asymmetric.
   
\section{Linear stability analysis}
\label{SecSability}
In this section, we discuss the linear stability of the self-gravitating kink solutions against small perturbations. This issue has been studied extensively in 5D brane world models~\cite{DeWolfeFreedmanGubserKarch2000,Giovannini2001a,Giovannini2002,Giovannini2003,ZhongLiu2013}, but remains untouched in the case of 2D.  The reducing of dimensions and the introducing of dilaton field make it impossible to analyze linear stability of 2D self-gravitating kinks by simply copying the stability analysis of 5D thick branes. For example, there are no vector and tensor perturbation in 2D, so the traditional scalar-vector-tensor decomposition~\cite{Giovannini2002,ZhongLiu2013} is no longer needed. Beside, in 2D there is no way to eliminate the non-minimal gravity-dilaton coupling by using conformal transformation.

It is convenient to discuss the linear stability in  the conformal flat coordinates
\begin{eqnarray}
\label{MetricRCoord}
ds^2=e^{2A(r)}\eta_{\mu\nu}dx^\mu dx^\nu,
\end{eqnarray}
where $r$ is defined through $dr\equiv e^{-A(x)}dx$. For simplicity, we use a prime and an overdot to represent the derivatives with respect to $r$ and $t$, respectively.

In this coordinates, the Einstein equations take the following form:
\begin{eqnarray}
\label{EqEinstrCoord}
\kappa  \phi '^2&=&4 A'\varphi '-2 \varphi ''-\varphi '^2,\\
\varphi '' &=& e^{2A} (\Lambda -\kappa  V).
\end{eqnarray}
The equation of motion for the scalar and dilaton fields are
\begin{eqnarray}
\phi ''&=&e^{2 A} \frac{dV}{d\phi},
\end{eqnarray}
and
\begin{eqnarray}
\label{eqVphiAr}
\varphi ''&=&2 A'',
\end{eqnarray}
respectively. Obviously, the general solution of Eq.~\eqref{eqVphiAr} is $\varphi=2A+\beta r+\varphi_0$, but as stated before, we will take $\beta=0=\varphi_0$. 

Equation \eqref{EqAphi} becomes
\begin{eqnarray}
2 A'''-4 A' A''+\kappa  \phi ' \phi ''=0,
\end{eqnarray} 
which, after integration, gives
\begin{eqnarray}
\label{EqAandphiPrim}
A''- {A'} ^2+\frac14 \kappa {\phi ' }^2 =0,
\end{eqnarray} 
where the integral constant has been taken as zero.

Now, let us consider small field perturbations around an arbitrary static background solution $\{\varphi(r), \phi(r), g_{\mu\nu}(r)\}$:
\begin{eqnarray}
\varphi(r)+\delta\varphi(r,t),\quad \phi(r)+\delta\phi(r,t),\quad g_{\mu\nu}(r)+\delta g_{\mu\nu}(r,t).
\end{eqnarray}
We also define
\begin{eqnarray}
\delta g_{\mu\nu}(r,t)\equiv e^{2A(r)} h_{\mu\nu}(r,t),
\end{eqnarray}
for convenience.

In the linear perturbation analysis of cosmological or brane world models, one usually decompose $h_{\mu\nu}$ into scalar, vector and tensor sectors~\cite{MukhanovFeldmanBrandenberger1992,KodamaSasaki1984}. Each sector can be discussed independently. In the present case, we have only one spatial dimension and no such decomposition is needed. So we will directly deal with the components of the metric perturbation
\begin{eqnarray}
h_{\mu\nu}=\left(
\begin{array}{cc}
 h_{00}(r,t) & \Phi (r,t) \\
 \Phi (r,t) & h_{rr}(r,t) \\
\end{array}
\right),
\end{eqnarray}
where we have renamed $h_{01}=h_{10}$ as $\Phi$, and $h_{11}$  as $h_{rr}$.

To the first order, the perturbation of the metric inverse is given by
\begin{eqnarray}
\delta g^{\mu \nu}=-e^{-2A} h^{\mu \nu}.
\end{eqnarray}
Note that the indices of $h$ are always raised or lowered with $\eta_{\mu\nu}$, thus, 
\begin{eqnarray}
h^{\mu \nu}\equiv \eta^{\mu\rho}\eta^{\nu\sigma}h_{\rho\sigma}=\left(
\begin{array}{cc}
 h_{00} & -\Phi \\
 -\Phi & h_{rr} \\
\end{array}
\right).
\end{eqnarray}

After linearization, the Einstein equations \eqref{eqEinstein} lead to three nontrivial perturbation equations, namely,  
the $(0,0)$ component: 
\begin{eqnarray}
\label{PertEq0}
&&2A'\delta \varphi '-2A'\varphi ' h_{rr}-2\delta \varphi ''-\delta \varphi '\varphi '+h_{rr}'\varphi '\nonumber\\
&+&2h_{rr}\varphi ''
+\frac{1}{2}h_{rr}\varphi '^2
=\kappa\left( \phi '\delta \phi '+ \phi ''\delta \phi -\frac{1}{2} \phi '^2 h_{rr}\right) ,
\end{eqnarray}
the $(0,1)$ or $(1,0)$ components:
\begin{eqnarray}
\label{PertEq1}
&&2A'\delta \varphi -2 \delta \varphi '- \varphi '\delta \varphi + \varphi '{h_{rr}}=\kappa \phi '\delta \phi,
\end{eqnarray}
and the $(1,1)$ component:
\begin{eqnarray}
\label{PertEq2}
2A'\delta \varphi '-2A'\varphi 'h_{rr}-\delta \varphi '\varphi '-2\ddot{\delta}\varphi +\frac{1}{2}h_{rr}\varphi '^2+\Xi \varphi '=\kappa \left( \phi '\delta \phi '-\phi ''\delta \phi -\frac{1}{2}\phi '^2h_{rr} \right).
\end{eqnarray}
Here we have defined a new variable $\Xi\equiv 2\dot{\Phi}-{h}_{00}'$.
One can testify that after using background equations \eqref{EqEinstrCoord}-\eqref{eqVphiAr}, Eq. \eqref{PertEq0} reduces to Eq. \eqref{PertEq1}.

Another independent equation comes from the perturbation of the scalar equation of motion:
\begin{eqnarray}
\label{PertEq3}
-\ddot{\delta}\phi +\delta \phi ''+2A'\frac{\phi ''}{\phi '}\delta \phi -\frac{\phi '''}{\phi '}\delta \phi -\frac{1}{2}\phi ' h_{rr}'-\phi ''h_{rr}+\frac{1}{2} \phi '\Xi=0.
\end{eqnarray}
One can also linearize the dilaton equation \eqref{eqDilatonxcoord}, but it does not offer new information further.

Therefore, we have three independent perturbation equations, i.e.,  \eqref{PertEq1}-\eqref{PertEq3}. But one should note that the perturbation variables are not all independent. The invariance of the dynamical equations under coordinate transformations
\begin{eqnarray}
x^\mu\to \tilde{x}^\mu=x^\mu+\xi^\mu(r, t)
\end{eqnarray} 
induces an invariance of the linear perturbation equations \eqref{PertEq1}-\eqref{PertEq3} under the following gauge transformations:
\begin{eqnarray}
\label{EqhmnTrans}
 \Delta h_{\mu\nu}  &\equiv& \widetilde{h}_{\mu\nu}-h_{\mu\nu} =-2 \xi_{(\mu, \nu)}
 -2 \eta_{\mu,\nu} A' \xi^{1} ,\\
 \Delta \delta \phi &\equiv& \widetilde{\delta \phi }-\delta \phi  = - \phi^{\prime}\xi^{1},\\
  \Delta \delta \varphi &\equiv& \widetilde{\delta \varphi}-\delta \varphi = - \varphi^{\prime} \xi^{1}.
  \end{eqnarray}
The components of $h_{\mu\nu}$ transform as
\begin{eqnarray}
\Delta h_{00}&=&2\partial _t\xi ^0+2A'  \xi ^1,
\\
\Delta \Phi&=&-\partial _t\xi ^1+\partial _r\xi ^0,
\\
\Delta h_{rr}&=&-2\partial _r \xi ^1-2 A' \xi ^1,
\end{eqnarray}
which means that the variable $\Xi= 2\dot{\Phi}-{h}_{00}'$ should transforms as 
\begin{eqnarray}
\Delta \Xi=-2\left[\ddot{\xi}^1+\left(A'\xi^1\right)'\right].
\end{eqnarray}
We see that the gauge degree of freedom $\xi^0$ has been canceled. 

  The residual gauge degree of freedom in $\xi^1$ allows us to eliminate one of the perturbation variables. Here we simply take $\delta\varphi=0$, with which Eq. \eqref{PertEq1} reduces to
 \begin{eqnarray}
\varphi '{h_{rr}}=\kappa \phi '\delta \phi,
\end{eqnarray}
and Eq. \eqref{PertEq2} becomes
\begin{eqnarray}
-2A'\varphi 'h_{rr}
+\frac{1}{2}h_{rr}\varphi '^2+\Xi \varphi '=\kappa \left( \phi '\delta \phi '-\phi ''\delta \phi -\frac{1}{2}\phi '^2h_{rr} \right).
\end{eqnarray}
  
  After eliminating $h_{rr}$ and $\Xi$, equation \eqref{PertEq3} can be written as a wave equation of $\delta\phi$:
\begin{eqnarray}
\label{eqDelPhi}
\ddot{\delta \phi }-\delta \phi ''+  V_{\text{eff}}(r)\delta \phi=0,
\end{eqnarray}
where the effective potential reads
\begin{eqnarray}
V_{\text{eff}}(r)=4A''-2A'\frac{\phi ''}{\phi '}-\varphi ''+2\left( \frac{\varphi ''}{\varphi '} \right) ^2-2\frac{\varphi '''}{\varphi '}+\frac{\phi '''}{\phi '}.
\end{eqnarray}
Using Eqs. \eqref{eqVphiAr}-\eqref{EqAandphiPrim}, one can obtain an useful identity:
 \begin{eqnarray}
\varphi''=\frac{\varphi'''}{\varphi'}+\frac{\phi''}{\phi'}\varphi'-2\frac{\phi''}{\phi'}\frac{\varphi''}{\varphi'},
\end{eqnarray}
which enable us to rewrite the effective potential as
\begin{eqnarray}
V_{\text{eff}}=\frac{\phi '''}{\phi '}
-2\frac{\phi ''}{\phi '}\frac{\varphi ''}{\varphi '}
+2\left( \frac{\varphi ''}{\varphi '} \right) ^2
-\frac{\varphi '''}{\varphi '},
\end{eqnarray}
or, in a more compact form
\begin{eqnarray}
V_{\text{eff}}=\frac{f ''}{f}, \quad \textrm{with} \quad f\equiv \frac{\phi '}{\varphi '}.
\end{eqnarray}
If we take $\delta\phi=\psi(r)e^{iwt}$, Eq.~\eqref{eqDelPhi} becomes a Schr\"odinger-like equation of $\psi(r)$:
\begin{eqnarray}
-\psi ''+  V_{\text{eff}}  \psi=w^2 \psi.
\end{eqnarray}
It is interesting to note that the Hamiltonian operator are factorizable:
\begin{eqnarray}
\hat{H}=-\frac{d^2}{dr^2}+V_{\text{eff}}=\hat{\mathcal{A}}\hat{\mathcal{A}}^\dagger,
\end{eqnarray}
with 
\begin{eqnarray}
\mathcal{A}=\frac{d}{d r}+\frac{{f}'}{f}, 
\quad \mathcal{A}^{\dagger}=-\frac{d}{d r}+\frac{{f}'}{f}.
\end{eqnarray}
According to the theory of supersymmetric quantum mechanics~\cite{CooperKhareSukhatme1995},  the eigenvalues of a factorizable Hamiltonian operator are semipositive definite, namely, $w^2\geq 0$. Therefore, static kink solutions are stable against linear perturbations. The zero mode ($w_0=0$) satisfies $ \mathcal{A}^{\dagger}\psi_0(r)=0$, and the solution reads
\begin{eqnarray}
\psi_0(r)\propto f=\frac{\phi '}{\varphi '}=\frac{\phi '}{2A '}.
\end{eqnarray}
Obviously, for any solution with a non-monotonic warp factor, $\psi_0(r)$ diverges at the extrema of $A$, and would be unnormalizable. Since it is not always possible to obtain the explicit expression of $x(r)$, it is useful  to transform $V_{\text{eff}}$  back to the $x$-coordinates:
\begin{eqnarray}
V_{\text{eff}}(x)&=&e^{2A}\left(A_x\frac{f_x}{f}+\frac{f_{xx}}{f}\right),
\end{eqnarray}
with $f(x)=\phi_x/\varphi_x$.

It should be note that the stability analysis presented so far are rather general and does not depend on the specific form of the solution, but  only on the general form of the metric \eqref{MetricRCoord} and of the action \eqref{action}.

Now, we move on to the specific solutions. For St\"otzel's sine-Gordon model and the polynomial model, the effective potentials read
\begin{eqnarray}
V_{\text{eff}}(x)&=&M^2 \cosh ^{-2 \kappa }(M x) \left[\kappa +2 \text{csch}^2(M x)+1\right],
\end{eqnarray}
and 
\begin{eqnarray}
V_{\text{eff}}(x)&=&
\frac{\exp\left[\frac{1}{12} \left(-6 c x+\text{sech}^2(x)-1\right)\right]}{{12 \sqrt[3]{\cosh (x)} \left[3 c+\tanh (x) \left(\text{sech}^2(x)+2\right)\right]^2}} \left\{-\text{sech}^2(x) \left[296\right.\right. \nonumber\\
&+&\left.\left. 702 c^2+\left(27 c^2-424\right) \text{sech}^2(x)+118 \text{sech}^4(x)
+\text{sech}^6(x)
+\text{sech}^8(x)
\right]\right. \nonumber\\
&+&\left.18 c \tanh (x) \left[3 c^2
+23 \text{sech}^4(x)-32 \text{sech}^2(x)+36\right]+540 c^2+208\right\},
\end{eqnarray}
respectively. For the later case, we have taken $\kappa=1$, for simplicity.

\begin{figure}[h]
\centering
\includegraphics[width=1\textwidth]{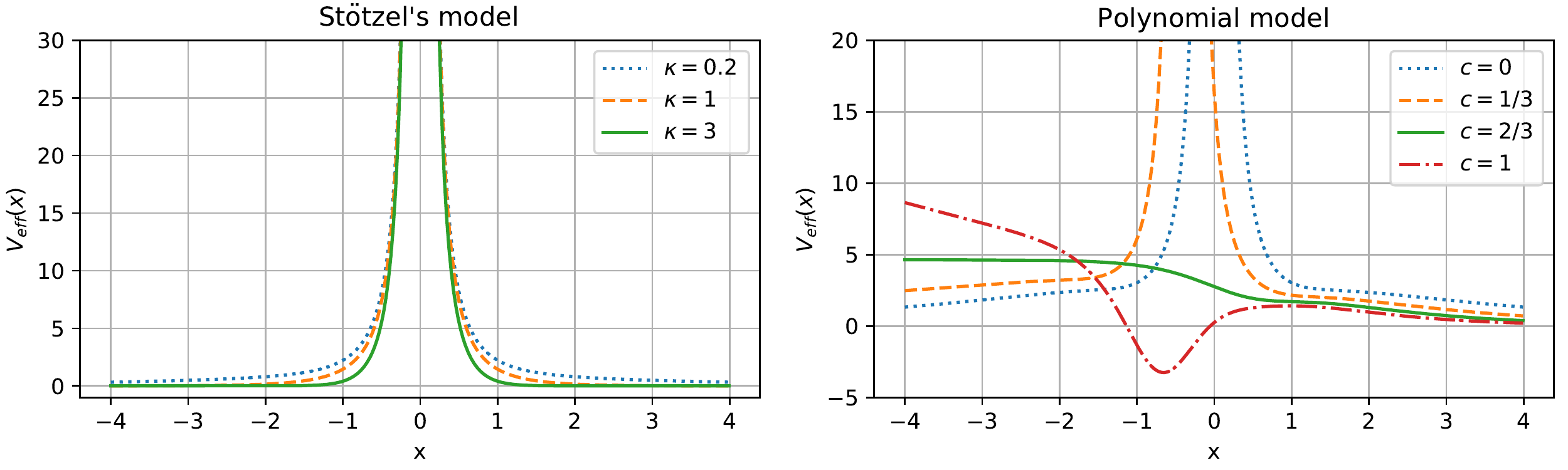}
\caption{Plots of $V_{\text{eff}}(x)$.  For polynomial model with $c=2/3$, $V_{\text{eff}}(x)$ becomes finite, and approaches to $4 \sqrt[3]{2} e^{-\frac{1}{12}}\approx 4.637$ as $x\to-\infty$.}
\label{figVeff}
\end{figure}

The profiles of the $V_{\text{eff}}(x)$ are depicted in Fig.~\ref{figVeff}. For St\"otzel's model, we take $m=\sqrt{2}$, $\Lambda=2\kappa$ such that $M\equiv \frac{1}{2} \sqrt{4 m^2-\frac{2\Lambda}{\kappa} }=1$, while keep $\kappa$ as a free parameter. We see that $V_{\text{eff}}$ is positive and divergent at $x=0$ for $\kappa=0.2$, 1 and 3. 

For the polynomial model, we take $c=0$, 1/3, 2/3 and 1 as examples. We see that $V_{\text{eff}}(x)$ diverges at $x=0$ for both $c=0$ and 1/3, while blows up at $x\to -\infty$ if $c=1$, but becomes finite when $c=2/3$. 

\begin{figure}[h]
\centering
\includegraphics[width=0.5\textwidth]{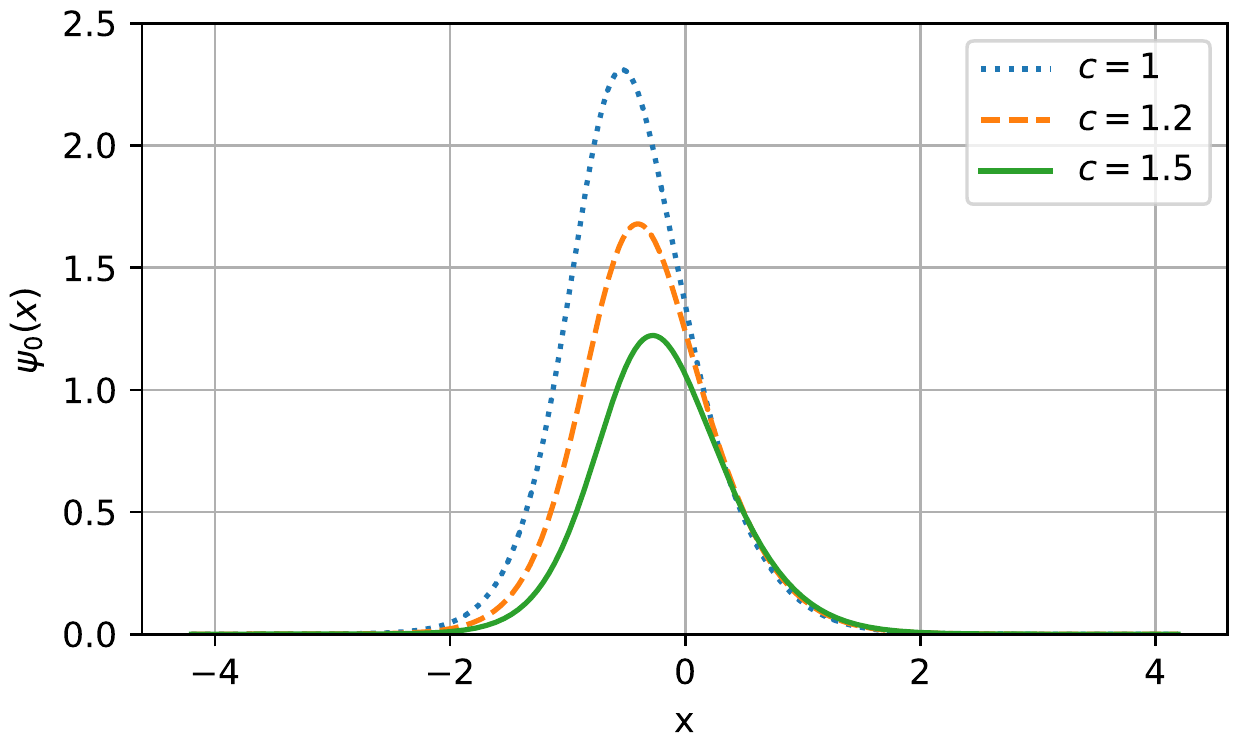}
\caption{Plots of $\psi_0(x)$ for the polynomial model with $\kappa=1$, $c=1$, 1.2 and 1.5.}
\label{FigZeroMode.pdf}
\end{figure}

It is worth to mention that in many 5D thick brane models the effective potentials of the scalar perturbation also have singularities, and the corresponding scalar zero modes are usually unnormalizable. Without normalizable scalar zero modes, these models are free of the problem of long range scalar fifth force~\cite{Giovannini2002,Giovannini2001a,ZhongLiu2013}. For the 2D self-gravitating kink solutions considered in this paper, however, we find an unusual situation where the zero mode might be normalizable, namely, the polynomial model with $c>2/3$. In this case, the zero mode reads
\begin{eqnarray}
\psi_0(x)=\mathcal{N}\frac{\phi_x}{2A_x}=-\mathcal{N}\frac{6 \text{sech}^2(x)}{3 c  +  \tanh (x) \left(\text{sech}^2(x)+2\right)},
\end{eqnarray}
where $\mathcal{N}$ is the normalization constant, and we have taken $\kappa=1$. The normalization of zero mode requires
\begin{eqnarray}
1&=&\int_{-\infty}^{+\infty} dr \psi_0^2(r)
=\mathcal{N}^2 \int_{-\infty}^{+\infty} dx e^{-A}  \left(\frac{\phi_x}{2A_x}\right)^2.
\end{eqnarray}
The integration can be done numerically, for instance, taking $c=1$, 1.2 and 1.5 we obtain $|\mathcal{N}|\approx$ 0.334, 0.446 and 0.598, respectively. Plots of $\psi_0(x)$ is depicted in Fig. \ref{FigZeroMode.pdf}.

\section{Summary and outlook}
\label{SecConc}

In this work, we revisited smooth self-gravitating kink solutions of a type of 2D dilaton gravity proposed by Mann et al.~\cite{MannMorsinkSikkemaSteele1991}. We first showed that exact kink solutions can be constructed with the aid of a first-order superpotential formalism \eqref{EqPhiW}-\eqref{eqAVarphi}  of the dynamical equations. This formalism has already been derived and used by St\"otzel in 1995, for 2D self-gravitating sine-Gordon model~\cite{Stoetzel1995}, but its virtue was not completely appreciated until the advent of 5D thick brane world models. After reproducing St\"otzel's solution~\cite{Stoetzel1995}, we reported another kink solution generated by a polynomial superpotential used in some 5D brane world models~\cite{EtoSakai2003,TakamizuMaeda2006,BazeiaMenezesRocha2014}. 

The main contribution of the present work, however, is a general analysis on the stability of static kink solutions under small linear perturbations. After eliminating the redundant gauge degrees of freedom, we derived a Schr\"odinger-like equation for the physical perturbation. We found that the Hamiltonian operator can be factorized as $\hat{H}=\hat{\mathcal{A}}\hat{\mathcal{A}}^\dagger$, which implies the stability of the solutions. Besides, the zero mode takes the form $\psi_0(r)\propto f\equiv\frac{\phi '}{\varphi '}=\frac{\phi '}{2A'}$, which diverges at the extrema of $A$.  For St\"otzel's model, the zero mode is  not normalizable, because the symmetric solution of the warp factor corresponds to a singularity of $\psi_0(r)$ at $r=0$. For the polynomial model, however, the zero mode is normalizable provides $c>2/3$.

It is natural to ask if the superpotential formalism and stability analysis of the present work can also be extended to other 2D dilaton gravity models, such as the CGHS model~\cite{CallanGiddingsHarveyStrominger1992} or other more general models~\cite{IkedaIzawa1993,TakahashiKobayashi2019}. The superpotential formalism also makes it possible discuss the application of the present model to the study of holographic RG flow~\cite{BianchiFreedmanSkenderis2001,KiritsisNittiSilvaPimenta2017}. We will leave these questions to our future works.
\section*{Acknowledgements}
 This work was supported by the National Natural Science Foundation of China (Grant Nos.~11847211, 11605127), Fundamental Research Funds for the Central Universities (Grant No.~xzy012019052), and China Postdoctoral Science Foundation (Grant No.~2016M592770).

\providecommand{\href}[2]{#2}\begingroup\raggedright\endgroup



\end{document}